\documentclass[apl,superscriptaddress,reprint]{revtex4-1}
\usepackage{amsmath}
\usepackage{graphicx}
\usepackage[T1]{fontenc}       
\usepackage{textcomp}
\usepackage{xcolor}
\usepackage{soul}

\begin{document}

\title{Probing electronic lifetimes and phonon anharmonicities in high-quality chemical vapor deposited graphene by magneto-Raman spectroscopy}

\author{ Christoph Neumann }
\email[Electronic address: ]{cneumann@physik.rwth-aachen.de}
\affiliation{ JARA-FIT and 2nd Institute of Physics, RWTH Aachen University, 52074 Aachen, Germany }
\affiliation{ Peter Gr\"unberg Institute (PGI-9), Forschungszentrum J\"ulich, 52425 J\"ulich, Germany }

\author{ Donatus Halpaap }
\affiliation{ JARA-FIT and 2nd Institute of Physics, RWTH Aachen University, 52074 Aachen, Germany }

\author{ Sven Reichardt }
\affiliation{ JARA-FIT and 2nd Institute of Physics, RWTH Aachen University, 52074 Aachen, Germany }
\affiliation{ Physics and Materials Science Research Unit, Universit\'e du Luxembourg, 1511 Luxembourg, Luxembourg }

\author{ Luca Banszerus }
\affiliation{ JARA-FIT and 2nd Institute of Physics, RWTH Aachen University, 52074 Aachen, Germany }

\author{ Michael Schmitz }
\affiliation{ JARA-FIT and 2nd Institute of Physics, RWTH Aachen University, 52074 Aachen, Germany }

\author{ Kenji Watanabe }
\affiliation{ National Institute for Materials Science, 1-1 Namiki, Tsukuba, 305-0044, Japan }

\author{ Takashi Taniguchi }
\affiliation{ National Institute for Materials Science, 1-1 Namiki, Tsukuba, 305-0044, Japan }

\author{ Bernd Beschoten }
\affiliation{ JARA-FIT and 2nd Institute of Physics, RWTH Aachen University, 52074 Aachen, Germany }

\author{ Christoph Stampfer }
\affiliation{ JARA-FIT and 2nd Institute of Physics, RWTH Aachen University, 52074 Aachen, Germany }
\affiliation{ Peter Gr\"unberg Institute (PGI-9), Forschungszentrum J\"ulich, 52425 J\"ulich, Germany }

\date{\today}


\begin{abstract}

We present a magneto-Raman study on high-quality single-layer graphene grown by chemical vapor deposition (CVD) that is fully encapsulated in hexagonal boron nitride by a dry transfer technique.
By analyzing the Raman D, G, and 2D~peaks, we find that the structural quality of the samples is comparable to state-of-the-art exfoliated graphene flakes.
From $B$-field dependent Raman measurements, we extract the broadening and associated lifetime of the G peak due to anharmonic effects.
Furthermore, we determine the decay width and lifetime of Landau level (LL) transitions from magneto-phonon resonances as a function of laser power.
At low laser power, we find a minimal decay width of 140~cm$^{-1}$ highlighting the high electronic quality of the CVD-grown graphene.
At higher laser power, we observe an increase of the LL decay width leading to a saturation, with the corresponding lifetime saturating at a minimal value of 18~fs.

\end{abstract}

\maketitle

Scalable fabrication methods for synthesizing high-quality graphene (Gr) samples are a major prerequisite for the integration of commercially competitive graphene-based devices.\cite{ferrari2015} 
Over the past years, various fabrication processes have been demonstrated to yield large single-crystal graphene.\cite{berger2006,sutter2008} 
Amongst others, chemical vapor deposition (CVD) has been shown to be a reliable technique for fabricating large graphene sheets with few lattice defects.\cite{li2009,bae2010,vlassiouk2011,petrone2012} 
Recently, wet chemistry-free transfer processes relying on mechanical delamination of CVD-grown graphene on copper were demonstrated to bring the fabrication of high-quality and low-cost graphene sheets for electronic and optoelectronic applications within reach.\cite{yoon2012,banszerus2015} 
In this work, we employ magneto-Raman spectroscopy to probe the quality of CVD graphene encapsulated in hexagonal boron nitride (hBN).
By investigating the resonant coupling of Landau level transitions to optical phonons, we find the strengths of these resonances to be comparable to those found in exfoliated graphene, demonstrating the high quality of our CVD graphene.
Moreover, we use a simple two-level system model to extract electronic and anharmonic broadenings from the most prominent resonance.
By varying the laser power, we find that the electronic lifetimes strongly depend on the laser power in the low-power regime, while they saturate to values of around 18~fs at higher laser power, which we associate with a saturation of the density of hot charge carriers in the system.

For our measurements, we use a low-temperature confocal Raman spectroscopy setup that is equipped with an x-y DC piezo-stage and a $\times$100 objective.
The sample is mounted in a liquid helium cryostat with a base temperature of 4.2~K.
In the cryostat, a magnetic field of up to 9~T can be applied perpendicular to the sample plane.
We use an excitation laser wavelength of 532~nm.
For detection, the light is guided via a single-mode fiber to a CCD-spectrometer with a 1200~lines/mm grating.

\begin{figure}
\centering
\includegraphics[draft=false,keepaspectratio=true,clip,width=1.0\linewidth]{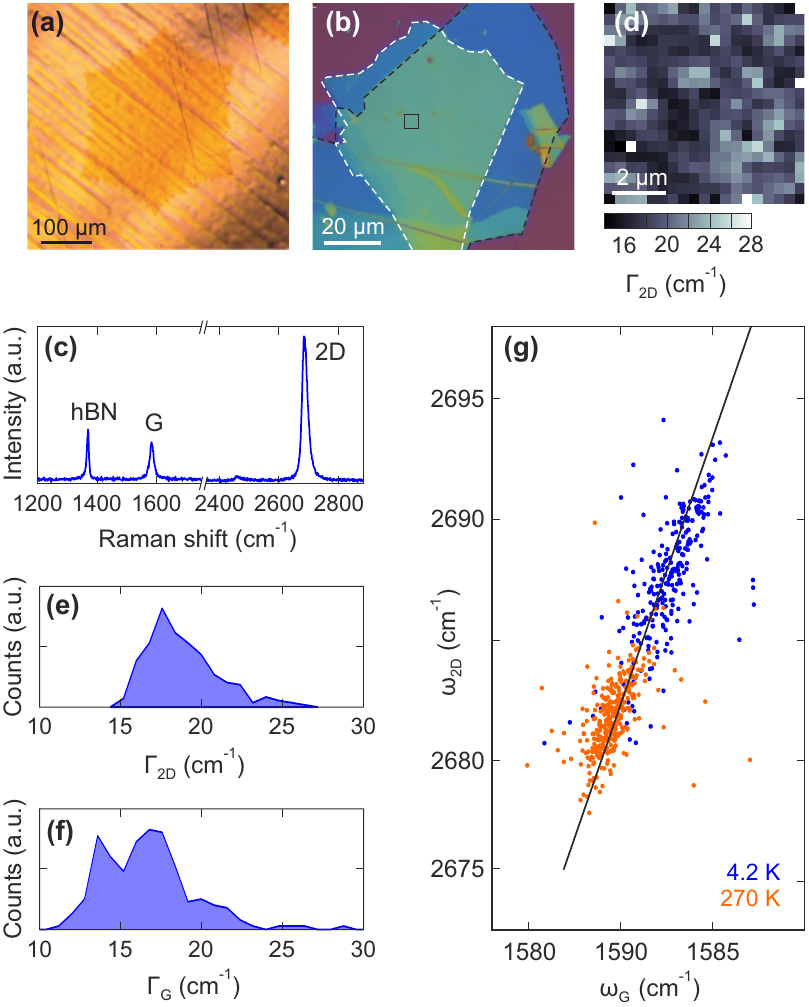}
\caption[FIG1]{
{ Sample characterization: }
(a) Optical image of a CVD single-layer graphene flake on a copper foil.
(b) Optical image of an hBN-Gr-hBN heterostructure. The black, dashed line outlines the boundaries of the top hBN-Gr stack, while the white, dashed line traces the contour of the bottom hBN flake. The black square indicates the position where the Raman map of panel (d) was obtained.
(c) Representative Raman spectrum taken at a temperature of 4.2~K. The characteristic hBN peak at around 1365~cm$^{-1}$,\cite{geick1966,nemanich1981} the Raman G~peak at around 1580~cm$^{-1}$, and the 2D~peak at around 2680~cm$^{-1}$ can be seen. The graphene D~peak, which is expected to appear at around 1345~cm$^{-1}$ in defected graphene is not observed.
(d) Scanning Raman image of $\Gamma_{2D}$. 
(e) Histogram of $\Gamma_{2D}$ at a temperature of 4.2~K.
(f) Histogram of $\Gamma_{G}$ at a temperature of 4.2~K.
(g) Scatter plot of $\omega_{2D}$ against $\omega_{G}$ from a Raman map at 4.2~K (blue) and 270~K (red). The black line indicates the slope of 2.2 due to strain induced shifts. All data in this figure are obtained with a laser power of $\sim$5~mW. }
\label{fig1}
\end{figure}

The single-layer graphene crystals are grown by low-pressure CVD on copper foils using CH$_4$/H$_2$.\cite{li2009,bae2010,vlassiouk2011,petrone2012} 
We employ a wet chemistry-free dry transfer process that yields high-quality CVD graphene devices, which consistently show low-temperature charge carrier mobilities on the order of 10$^5$~cm$^2$/Vs and doping values below 2 $\times$ 10$^{11}$~cm$^{-2}$ (for more details please see Reference~\onlinecite{banszerus2015}).

In this approach, the CVD graphene sheet is picked up from the copper foil with a multilayer hBN flake only by van der Waals adhesion.
In a next step, the hBN-Gr stack is placed onto a second hBN crystal located on a SiO$_2$/Si$^{++}$ wafer.
As single-layer CVD graphene flakes on copper can be grown on length scales of up to $10^{-4} - 10^{-3}$~m (Figure~1a), remarkably large hBN-Gr-hBN stacks can be made with this technique.
Currently the size of the samples is only limited by the dimensions of the exfoliated hBN crystals.
As an example, a $50 \times 40$ \textmu m$^2$ hBN-Gr-hBN heterostructure is shown in Figure~1b.

To probe the quality of the sample by magneto-Raman spectroscopy, a resonant coupling of Landau level (LL) transitions to the Raman G~peak needs to be observable.
For this, the sample has to meet two criteria:
First, a high electronic quality is needed to ensure sufficiently long lifetimes of the Landau level excitations.
Second, the specimen must have a low charge carrier density (ideally below $3 \times 10^{11}$~cm$^{-2}$).\cite{neumann2015} 
Otherwise the coupling of the LL transitions starts to be suppressed due to the Pauli principle (i.e. Pauli blocking).

We obtain information on the amount of lattice defects and the structural quality of the sample from scanning, confocal Raman spectroscopy.
A representative Raman spectrum of the investigated CVD sample is shown in Figure~1c.
We do not find a noticeable Raman D~peak on the entire sample, suggesting a very low amount of lattice defects. \cite{ferrari2007,canccado2011} 
In Figure~1d, a Raman map displaying the full width at half maximum (FWHM) of the 2D~peak, $\Gamma_{2D}$, is shown.
Within the entire sample, we find a small line width, indicating the absence of large folds and bubbles.\cite{neumann2015b} 
In general, the values of $\Gamma_{2D}$ are mostly between 16~cm$^{-1}$ and 22~cm$^{-1}$ (see Figure~1e), demonstrating the low amount of strain variations on length scales below the laser spot size (i.e. on the nm-scale),\cite{neumann2015b} which has recently been shown to be a precondition for high-mobility graphene devices.\cite{couto2014} 

Information on charge carrier doping, on the other hand, can be obtained from the Raman G~peak.
The FWHM of the G~peak, $\Gamma_{G}$, across the sample is mostly between 12~cm$^{-1}$ and 22~cm$^{-1}$,  
indicating small variations of the generally low charge carrier doping
of the CVD sample (Figure~1f).\cite{pisana2007,stampfer2007} 
The homogeneity of the charge carrier doping in the device is further highlighted by the linear dependence of the frequencies of the G and 2D~peaks in the scatter plot shown in Figure~1g.
The data points follow a straight line with a slope of 2.2, which coincides with the expected relative shifts of the Raman modes for mechanical strain.\cite{lee2012} 
Furthermore, the spread of the data points around the 2.2 axis is low.
Consequently, differences in $\omega_{G}$ and $\omega_{2D}$ along the sample can be attributed to differences in the local strain and do not originate from variations of the local charge carrier doping.

\begin{figure}
\centering
\includegraphics[draft=false,keepaspectratio=true,clip,width=1.0\linewidth]{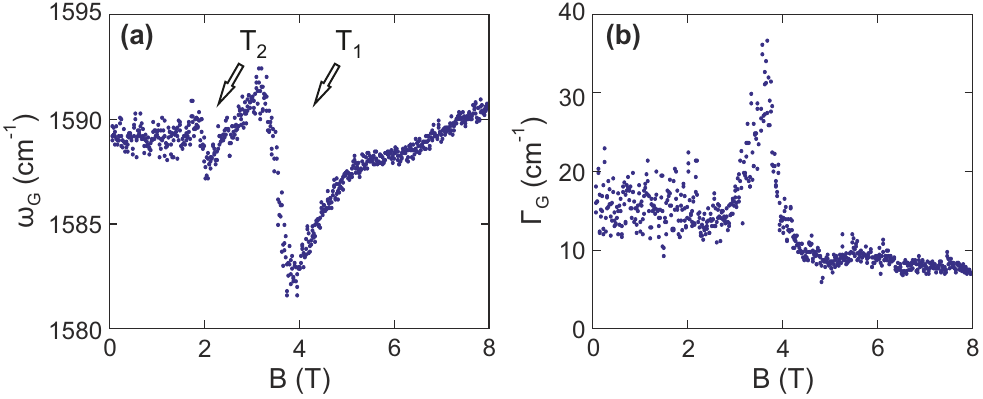}
\caption[FIG2]{
{ Magneto-Raman spectroscopy: }
(a) $\omega_G$ as obtained from single Lorentzian fits to each individual spectrum as a function of magnetic field. The arrows point out the $T_1$ and $T_2$ MPRs. The data is obtained at a temperature of 4.2~K with a laser power of $\sim$5~mW.
(b) Corresponding plot of $\Gamma_G$ versus magnetic field. }
\label{fig2}
\end{figure}

As the two required conditions are fulfilled, we perform magnetic field-dependent Raman spectroscopy measurements.
In Figures~2a and 2b we show the frequency of the G~peak, $\omega_G$, and the broadening, $\Gamma_G$, respectively, as obtained from each spectrum by single Lorentzian fits as a function of magnetic field.
We observe the typical resonant behavior of the G~peak \cite{neumann2015} in both $\omega_G$ and $\Gamma_G$ at $B = 1.99$~T and at $B = 3.56$~T, which can be attributed to the so-called $T_2$ and $T_1$ magneto-phonon resonances (MPRs), respectively.
This phenomenon can be well understood by considering the interaction of the phonon mode with Landau level transitions.\cite{ando2007} 
If the energy of a LL transition matches the energy of the phonon without magnetic field, i.e. if $T_n = \hbar \omega_{\mathrm{ph}}$, an MPR occurs.
Here, $T_n = \varepsilon_n + \varepsilon_{n+1}$, with $\varepsilon_n = v_\mathrm{F}\sqrt{2 e \hbar B n}$, represents the energy of the $T_n$ LL transition and $\hbar \omega_{\mathrm{ph}} \approx 196$~meV.

By investigating eight different measurements (with different laser powers; see Fig. 3) and taking the mean $B$ field values at which the $T_1$ and $T_2$ magneto-phonon resonances appear we deduce effective Fermi velocities $v_{\mathrm{F},T_1} = 1.19 \pm 0.01 \times 10^6$~m/s and $v_{\mathrm{F},T_2} = 1.23 \pm 0.03 \times 10^6$~m/s that depend on the exact transition as predicted by theory.\cite{iyengar2007,shizuya2010,shizuya2011,chizhova2015} 
Just as in high-quality hBN-encapsulated exfoliated graphene, these values vary from the usual single-particle picture Fermi velocity of $\sim1.06 \times 10^6$~m/s, as e.g. found by magneto-Raman measurements for graphene on graphite.\cite{yan2010,faugeras2011}

\begin{figure}
\centering
\includegraphics[draft=false,keepaspectratio=true,clip,width=1.0\linewidth]{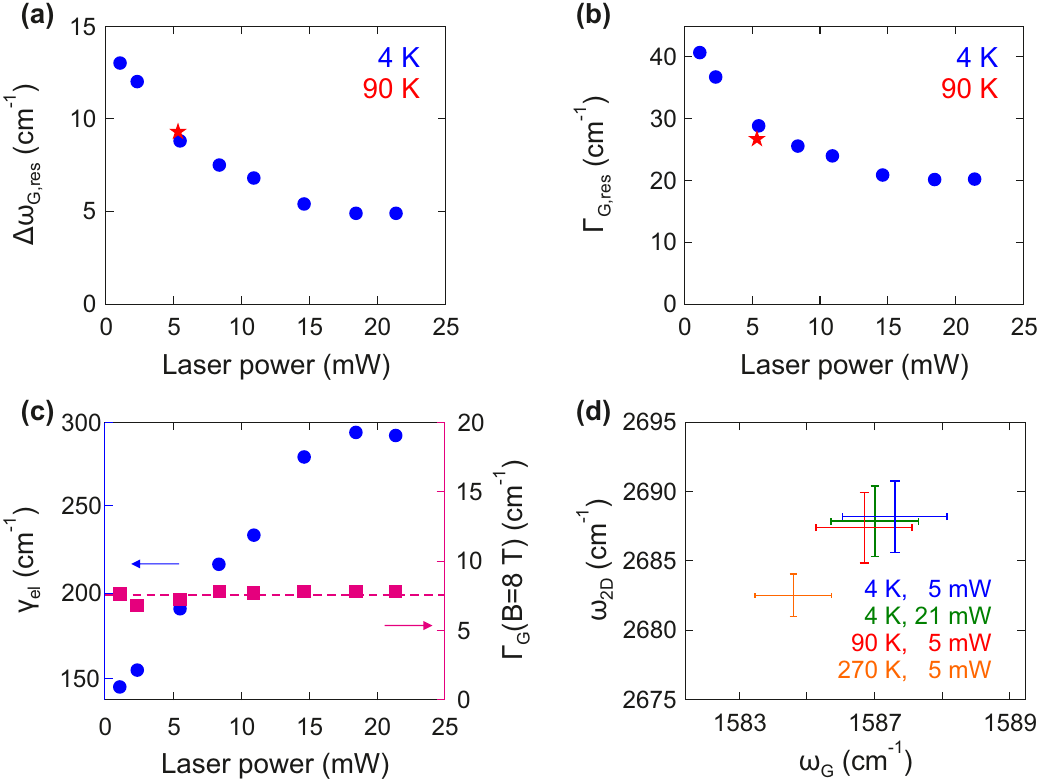}
\caption[FIG3]{
{ Resonance strength: }
(a) Maximum shift of the G~peak around the $T_1$ resonance, $\Delta\omega_{G,\mathrm{res}}$, against laser power. The blue data points were taken at a temperature of 
4.2~K, while the red star was obtained at 90~K.
(b) Corresponding extracted maximum broadening of the G~peak at the $T_1$ resonance, $\Gamma_{G,\mathrm{res}}$, against laser power.
(c) $\Gamma_G$ at $B=8$~T as a function of laser power (purple squares). The dashed line represents the average value used as a measure for $\gamma_{\mathrm{ph}}$. The blue data points represent $\gamma_{\mathrm{el}}$ as calculated from $\Gamma_{G,\mathrm{res}}$ using Eq.~2.
(d) Mean values (and standard deviations) of $\omega_{2D}$ and $\omega_{G}$ from four different Raman maps. The specific parameters used for the maps are given in the figure. }
\label{fig3}
\end{figure}

Next we investigate the strength of the MPRs, focusing on the most prominent $T_1$-MPR.
At low laser power ($\sim$1~mW), we find a maximum shift of the G~peak around the resonance of $\Delta\omega_{G,\mathrm{res}} = 13$~cm$^{-1}$ and a maximum value of $\Gamma_{G}$ at the resonance of $\Gamma_{G,\mathrm{res}} = 41$~cm$^{-1}$, which is in excellent agreement with the findings for hBN-encapsulated, exfoliated graphene discussed in Reference~\onlinecite{neumann2015}.
This is good evidence for the low amount of disorder in our CVD sample.
However, as evident from Figures~3a and 3b, both $\Delta\omega_{G,\mathrm{res}}$ and $\Gamma_{G,\mathrm{res}}$ significantly decrease with increasing laser power.
To better understand the underlying physical origin behind this decrease, we study the behavior of the $T_1$~resonance in more detail.
Near the $T_1$~resonance, the MPR can be treated as a coupled two-level system, with the two levels being the G~mode phonon and the Landau level excitation $T_1$.

The energies of the coupled modes are given by
\begin{equation}
\begin{split}
\hbar\omega_{\pm}=&\frac{\hbar\omega_{\mathrm{ph}}+T_{1}}{2}-\frac{i}{2}\frac{\gamma_{\mathrm{ph}}+\gamma_{\mathrm{el}}}{2} \\
&\pm\frac{1}{2}\sqrt{\left(\hbar\omega_{\mathrm{ph}}-T_{1}-\frac{i}{2}(\gamma_{\mathrm{el}}-\gamma_{\mathrm{ph}})\right)^{2}+4g^{2}},
\end{split}
\end{equation}
where $g = \sqrt{\lambda/2}\hbar\omega_B$ is the $B$~field dependent electron-phonon coupling constant, in which $\omega_B = v_{\mathrm{F}} \sqrt{2 e B / \hbar}$ denotes the cyclotron frequency, $\lambda \approx 4 \times 10^{-3}$ is the dimensionless electron-phonon coupling constant \cite{faugeras2009,yan2010,kim2013,neumann2015,ando2007,kuhne2012} and $\gamma_{\mathrm{ph}}$ and $\gamma_{\mathrm{el}}$ are the decay widths of the phonon and $T_1$ LL excitation, respectively.
Note that due to the photo-induced doping effect occurring in these structures,\cite{ju2014,neumann2015} the Fermi level is pinned close to the charge neutrality point and hence $g$ does not depend on the charge carrier concentration.
Within the framework of this model, the strength of the resonance, which is characterized by the value of $\Gamma_{G,\mathrm{res}} = -2\,\mathrm{Im}(\omega_{+}(B_{\mathrm{res}}))$, is defined by three parameters: the electron-phonon coupling constant $\lambda$, the decay width of the phonon mode due to non-electronic decay channels $\gamma_{\mathrm{ph}}$ and the decay width of the LL excitation $\gamma_{\mathrm{el}}$.
The electron-phonon coupling constant $\lambda$ has been determined in a variety of studies, where very similar values were found in all cases,\cite{faugeras2009,yan2010,kim2013,neumann2015,kuhne2012} indicating that it is very robust to changes of the laser power and even the choice of substrate material and electronic quality of the investigated sample.
To determine a value for the anharmonic phonon broadening $\gamma_{\mathrm{ph}}$, we measure the FWHM of the G~peak at 8~T.
In this case, the coupling of LL transitions to the G~mode is highly suppressed since no LL transition is energetically close to the phonon mode.\cite{neumann2015,neumann2015b} The width of the G~peak is therefore devoid of any influences of the electronic system and only anharmonic broadening effects and possibly contributions from averaging over strain variations within the laser spot \cite{neumann2015b} remain.
As shown in Figure~3c (purple squares), $\Gamma_G$ at 8~T is very robust with respect to changes of the laser power, which supports the absence of electronic broadening effects, since changing the laser power primarily modifies the electronic system as shown below. 
By averaging $\Gamma_G$($B$=8~T) measured at different laser powers, we obtain an estimate for $\gamma_{\mathrm{ph}}$ of 7.6~cm$^{-1}$ (see dashed line in Figure~3c), which corresponds to a lifetime of $\tau_{\mathrm{ph}} = \hbar/\gamma_{\mathrm{ph}} \approx 708$~fs. Note that this value of $\Gamma_G$ at 8~T is slightly higher than the smallest values of around 5~cm$^{-1}$ measured in previous studies on exfoliated graphene \cite{neumann2015,neumann2015b} and also higher than theoretical predictions from ab inito methods.\cite{paulatto2013} 
Consequently, it is likely that averaging effects over strain variations make a contribution to the observed $\Gamma_G$ at 8~T.\cite{neumann2015b} In this case the actual $\gamma_{\mathrm{ph}}$ would be smaller than 7.6~cm$^{-1}$.
The value for $\gamma_{\mathrm{ph}}$ can further be used to determine the decay width of the $T_1$ LL transition.
Inverting Eq.~1 yields a simple expression for $\gamma_{\mathrm{el}}$ at the resonance:
\begin{equation}
\gamma_{\mathrm{el}}=\Gamma_{G,\mathrm{res}}+\frac{4g_{\mathrm{res}}^{2}}{\Gamma_{G,\mathrm{res}}-\gamma_{\mathrm{ph}}},
\end{equation}
where the coupling constant at the resonance is given by $g_{\mathrm{res}} = \sqrt{\lambda/2}\hbar\omega_{\mathrm{ph}}/(\sqrt{2}+1)$.
The value of $\gamma_{\mathrm{el}}$ can thus be determined by measuring the width of the G~peak at the $T_1$~resonance.
In Figure~3c, we show the extracted values of $\gamma_{\mathrm{el}}$ as a function of laser power.
In the investigated range, we find an increase of $\gamma_{\mathrm{el}}$ by a factor of around 2.
Furthermore, we observe a saturation of $\gamma_{\mathrm{el}}$ in agreement with the saturation of $\Delta\omega_{G,\mathrm{res}}$ and $\Gamma_{G,\mathrm{res}}$.
The saturation value of $\gamma_{\mathrm{el,sat}} \approx 293 $~cm$^{-1}$ corresponds to a lifetime of the $T_1$ transition of $\tau_{\mathrm{el,sat}} \approx 18$~fs.
This dependence of $\gamma_{\mathrm{el}}$ on the laser power cannot be explained by an increase of the lattice temperature when increasing the laser power as illustrated by the red stars in Figures~3a and 3b, which represent an MPR measurement taken at a temperature of 90~K and a laser power of 5~mW.
Since the MPR is equally strong at 4~K, the strength of the resonance remains unaffected by a change of the lattice temperature.
Moreover, Figure~3d shows that even at maximum laser power (21.4~mW) we locally heat the sample up to less than 90~K, since the average shifts of $\omega_G$ and $\omega_{2D}$ for 21.4~mW at 4.2~K are smaller than the ones for 5~mW at 90~K.
We suggest that the increase of the decay width of the $T_1$ Landau level excitation with laser power is due to the increasing number of optically excited hot charge carriers in the system, leading to increased scattering rates and hence reduced lifetimes of LL excitations.
Since the width of the G~peak at zero magnetic field does not change with laser power, we can rule out effective changes of the Fermi level, which would also lead to a narrowing of the G~peak.

In summary, we investigated Raman spectra of high-quality CVD-grown graphene samples encapsulated in hBN with and without a magnetic field.
In particular, the small FWHM of the Raman 2D~peak between 16 cm$^{-1}$ and 22 cm$^{-1}$ indicates a similarly high quality and flatness of the CVD graphene sample as compared to state-of-the-art exfoliated graphene.
By applying a magnetic field perpendicular to the sample, we were able to observe magneto-phonon resonances at low $B$~fields.
We extracted effective Fermi velocities for both the $T_1$- and $T_2$-MPRs of $v_{\mathrm{F},T_1} = 1.19 \pm 0.01 \times 10^6$~m/s and $v_{\mathrm{F},T_2} = 1.23 \pm 0.03 \times 10^6$~m/s.
The strong deviation from the Fermi velocity valid within a single-particle picture in graphene ($1.06 \times 10^6$~m/s) shows the importance of many-body interaction effects for the correct description of the electronic spectrum of high-quality CVD graphene samples.
From the width of the G~peak at the $T_1$-MPR, we extracted the lifetime of the LL transition as a function of laser power, which is comparable to recent findings in encapsulated exfoliated graphene flakes \cite{neumann2015} demonstrating the high electronic quality of our samples.
By varying the laser power, we further observed a strong decrease of the electronic lifetimes with increasing laser power, reaching a saturation value of around 18~fs.
We attribute this effect to the saturation of the optical excitation of charge carriers. At the same time, the anharmonic contribution to the broadening of the G peak did not change with increasing laser power.

We gratefully acknowledge support by the Helmholtz Nanoelectronic Facility (HNF), the DFG, the ERC (GA-Nr. 280140), and the EU project Graphene Flagship (contract no. NECT-ICT-604391). S.R. acknowledges funding by the National Research Fund (FNR) Luxembourg.

\phantomsection
\providecommand*\mcitethebibliography{\thebibliography}
\csname @ifundefined\endcsname{endmcitethebibliography}
  {\let\endmcitethebibliography\endthebibliography}{}

\end{document}